\documentclass{article}

\usepackage{latexsym}

\usepackage{amssymb}
 \thinmuskip =
0.5\thinmuskip \medmuskip = 0.5\medmuskip \thickmuskip =
0.5\thickmuskip \arraycolsep = 0.3\arraycolsep

\newcommand{\R}{\mathbb{R}}

\def\prfe{\hspace*{\fill} $\Box$

\smallskip \noindent}
\newtheorem{Theorem}{Theorem}
\newtheorem{Proposition}{Proposition}
\newtheorem{Lemma}{Lemma}

\DeclareFontFamily{OT1}{rsfs}{}
\DeclareFontShape{OT1}{rsfs}{m}{n}{ <-7> rsfs5 <7-10> rsfs7
<10->rsfs10}{} \DeclareMathAlphabet{\mycal}{OT1}{rsfs}{m}{n}

\title{Global classical solutions to the 3D Nordstr\"om-Vlasov system}
\author{Simone Calogero\\[0.3cm]
Institutt for matematiske fag\\
NTNU Alfred Getz' vei 1 N-7491\\
Trondheim, Norway\\
E-mail: simone.calogero@math.ntnu.no}
\date { }
\begin{document}
\maketitle
\begin{abstract}
The Nordstr\"om-Vlasov system describes the evolution of
self-gravitating collisionless matter in the framework of a
relativistic scalar theory of gravitation. We prove global
existence and uniqueness of classical solutions for the
corresponding initial value problem in three dimensions when the
initial data for the scalar field are smooth and the initial
particle density is smooth with compact support.
\end{abstract}
\section{Introduction}\label{intro}
\setcounter{equation}{0} This paper is concerned with the Cauchy
problem for the Nordstr\"om-Vlasov system. The latter is a Lorentz
invariant kinetic model describing the evolution of
self-gravitating collisionless matter under the assumption that
the gravitational forces are mediated by a scalar field. In a
system of Cartesian coordinates $(t,x)$, $t\in\R,\,x\in\R^3$, the
Nordstr\"om-Vlasov system is given by
\begin{equation}\label{wavet}
\partial_t^2\phi-\Delta\phi=-\mu,
\end{equation}
\begin{equation} \label{mudeft}
\mu(t,x) = \int f(t,x,p)\,\frac{dp}{\sqrt{1+|p|^2}},
\end{equation}
\begin{equation} \label{vlasovt}
S f - \left[(S\phi)\,p + (1+|p|^2)^{-1/2} \nabla_x\phi
\right]\cdot\nabla_p f = 4 f\, S\phi .
\end{equation}
Here $p \in \R^3$ is the momentum variable, $f=f(t,x,p)$ is the
particle density in phase-space, $\phi=\phi(t,x)$ is the scalar
gravitational field generated by the particles and
\[
S =\partial_t+\widehat{p}\cdot\nabla_x,\quad \widehat{p} =
\frac{p}{\sqrt{1+|p|^2}};
\]
$S$ is the free-transport operator, $\widehat{p}$ denotes the
relativistic velocity of a particle with momentum $p$. Units are
chosen such that the mass of each particle, the gravitational
constant and the speed of light are equal to unity.  A solution
$(f,\phi)$ of this system is interpreted as follows: The
space-time is a four-dimensional Lorentzian manifold with a
conformally flat metric which, in the coordinates $(t,x)$, takes
the form
\begin{equation}\label{metric}
g_{\mu\nu}=e^{2\phi} \textrm{diag}(-1,1,1,1),\ \mu,\nu=0,\dots 3.
\end{equation}
The particle density on the mass shell in this metric is
$\mathfrak{f}=e^{-4\phi}f(t,x,e^\phi p)$, but it is more
convenient to work with $f$ and $\phi$ as the dynamic variables.
The Vlasov equation (\ref{vlasovt}) is equivalent to the condition
that $\mathfrak{f}$ is constant along the geodesics flow of the
metric (\ref{metric}). The right hand side of the field equation
(\ref{wavet}) is the trace of the stress energy tensor associated
to $f$ with respect to the background Minkowski metric. We refer
to \cite{C} for a derivation of the equations.

Although scalar theories of gravity are not physically correct,
they may provide useful simplified models for General Relativity
\cite{ST}. Moreover, scalar fields play a central role in modern
theories of classical and quantum gravity \cite{DEF}. The
physically correct relativistic model for self-gravitating
collisionless matter is the Einstein-Vlasov system, which is
discussed for instance in \cite{And,RR}.  Due the very complicated
structure of the Einstein equations, the evolution problem for the
Einstein-Vlasov system remains poorly understood, even in
spherical symmetry. In fact, global existence and uniqueness of
(asymptotically flat) solutions to the Einstein-Vlasov system has
been proved only for small data in spherical symmetry \cite{RR}.
As opposed to this, the Cauchy problem for the Vlasov-Poisson
system, which is the non-relativistic limit of the Einstein-Vlasov
system \cite{RR2, Re}, is by now well-understood, cf. \cite{LP,
Pf, R, Sch}. In \cite{CL} it is shown that Vlasov-Poisson is the
non-relativistic limit of the Nordstr\"om-Vlasov system as well.
However, global existence of classical solutions for the
Nordstr\"om-Vlasov system and related models---such as the
Vlasov-Maxwell system of plasma physics---has not yet been
established. In this paper we show that this fundamental question
has a positive answer for the Nordstr\"om-Vlasov system.
Precisely, we shall prove the following
\begin{Theorem}\label{main}
Given $f_0:\R^3\times\R^3\to [0,\infty)$ and
$\phi_0,\phi_1:\R^3\to \R$ satisfying
\[
f_0\in C^1_c,\ \phi_0\in C^3_b\cap H^1,\ \phi_1\in C^2_b\cap L^2,
\]
there exists a unique global solution $(f,\phi)$ of the
Nordstr\"om-Vlasov system in the class
\[
(f,\phi)\in C^1([0,\infty)\times\R^3\times\R^3)\times
C^2([0,\infty)\times\R^3),
\]
such that $(f,\phi)_{|t=0}=(f_0,\phi_0)$ and
$(\partial_t\phi)_{|t=0}=\phi_1$.
\end{Theorem}

The proof of Theorem \ref{main} relies upon three main tools. A
first one is the representation formula for the time derivative of
the field given in \cite[Prop.~1]{CR1}. It turns out that
estimating $\partial_t\phi$ is enough to obtain global existence.
A second important tool is given by the (null cone) energy
estimates which derive from the energy identity
\begin{equation}\label{localcon}
\partial_te + \nabla_x\cdot\mathfrak{p}=0,
\end{equation}
where
\[
e(t,x) = \int \sqrt{1+|p|^2}\,f\,dp +
\frac{1}{2}(\partial_t\phi)^2 + \frac{1}{2}(\nabla_x\phi)^2, \quad
\mathfrak{p}(t,x) = \int p\,f\,dp -\partial_t\phi\nabla_x\phi.
\]
Upon integration, (\ref{localcon}) leads to the identities
\begin{equation}\label{encons}
\int\int \sqrt{1+|p|^2}\,f\,dp\,dx +
\int\left[\frac{1}{2}(\partial_t\phi)^2 +
\frac{1}{2}(\nabla_x\phi)^2\right]\,dx=\rm{const.},
\end{equation}
\begin{equation}\label{nullcone}
\int_{|x-y|\leqslant
t}(e+\mathfrak{p}\cdot\omega)(t-|x-y|,y)\,dy=\int_{|x-y|=
t}e(0,y)\,dS_y,\quad\omega=\frac{(y-x)}{|x-y|}.
\end{equation}
Note also that
\[
e+\mathfrak{p}\cdot\omega=\int(\sqrt{1+|p|^2}+\omega\cdot p)\,f\,dp+\frac{1}{2}(\omega\wedge\nabla_x\phi)^2+\frac{1}{2}(\partial_t\phi-\omega\cdot\nabla_x\phi)^2.
\]
We shall refer to (\ref{nullcone}) as the null cone energy
identity, while (\ref{encons}) is the usual conservation of total
energy. They imply
\begin{equation}\label{energyestimate}
\|\partial_t\phi(t)\|_{L^2(\R^3)}+\|\nabla_x\phi(t)\|_{L^2(\R^3)}\leqslant
{\rm const.},
\end{equation}
\begin{equation}\label{nullconeenergyestimate}
\|(\omega\wedge\nabla_x\phi)(t,x)\|_{L^2(\Lambda_{t,x})}+\|(\partial_t\phi-\omega\cdot\nabla_x\phi)(t,x)\|_{L^2(\Lambda_{t,x})}\leqslant
{\rm const.}\times t^2,
\end{equation}
where $\R^4\supset\Lambda_{t,x}=\{(t-|x-y|,y):\,|x-y|\leqslant
t\}_{y\in\R^3}$ is the past light cone with vertex at $(t,x)$ and
base on $t=0$. The energy estimates are used to bound
$\partial_t\phi$. For this purpose the representation formula for
$\partial_t\phi$ must be rewritten in a proper way to single out
the contributions which are bounded by the null cone energy. All
terms in the integral representation of $\partial_t\phi$ but one
can be estimated using the energy estimates. The remaining term is
estimated using the third---and most important---main ingredient
in the proof of Theorem \ref{main}, which is the $L^2$ version of
a lemma due to C.~Pallard, see \cite[Lemma~1.1]{Pa}. This crucial
lemma establishes an $L^\infty$ bound for the time integral of
functions evaluated along the characteristics of the Vlasov
equation. As we shall need a slightly different formulation of the
result proved in \cite{Pa}, a proof thereof will be given when it
is needed.

We remark that prior to the present result, global existence
theorems for Nordstr\"om-Vlasov have been proved under certain
restrictions, such us small data \cite{F}, spherical symmetry
\cite{ACR} and for the 2-dimensional system \cite{L}. Global
existence of weak solutions is established in \cite{CR2}. The
proofs of these results make use of techniques originally
developed to study the Cauchy problem for the Vlasov-Maxwell
system, see \cite{BGP, DipLio, GS, GS2, GSh2, KS, Rein}. This
suggests that if the analogue of Theorem \ref{main} hold for the
Vlasov-Maxwell system, the proof thereof might rely upon similar
arguments as to the ones presented in this paper. However, since
for Vlasov-Maxwell one has to estimate a vector field instead of a
scalar field, the proof of global existence in the plasma physics
case might be considerably more difficult and require some
additional non-trivial idea.

\section{Preliminaries} Throughout the paper we denote by $C(t)$
any continuous non-decreasing positive function of time. If it is
a constant, we denote it simply by $C$. The characteristics of the
differential operator in the left hand side of (\ref{vlasovt}) are
the solutions of
\begin{equation}
\dot{x}=\widehat{p},\quad \dot{p}=-(S\phi)\,p - (1+|p|^2)^{-1/2}
\nabla_x\phi\label{char}
\end{equation}
and we denote by $(X,P)(s)$ the characteristic curve satisfying
$(X,P)(t)=(x,p)$. Note that $(X,P)(s)$ also depends on $(x,p)$,
but this is not reflected in our notation. The function
$e^{-4\phi}f$ is constant along the solutions of (\ref{char}). We
deduce that
\begin{equation}\label{reprf}
f(t,x,p)=f_0(X(0),P(0))\exp\left[4\phi(t,x)-4\phi_0(X(0))\right],
\end{equation}
whence
\begin{equation}\label{estf}
\|e^{-4\phi}f(t)\|_\infty\leqslant C.
\end{equation}
Let $\phi=\phi_{\mathrm{hom}}+\psi$, where $\psi$ is the solution
of (\ref{wavet})  with zero data and $\phi_\mathrm{hom}$ solves
$\Box\phi=0$ with data $(\phi_0,\phi_1)$. Since $f\geqslant 0$,
then $\psi\leqslant 0$. Therefore
\begin{equation}\label{est1}
\phi(t,x)\leqslant C(t),
\end{equation}
and
\begin{equation}\label{est2}
e^{\phi}|\phi|\leqslant C(t).
\end{equation}
For (\ref{est2}) we used that $\sup_{\xi\leqslant
0}\left(e^\xi|\xi|\right)\leqslant C$. Note also that
$\|f(t)\|_\infty\leqslant C(t)$. In \cite{CR1} it is proved that
the Cauchy problem for the Nordstr\"om-Vlasov system has a unique
classical solution locally in time. Let $T_\mathrm{max}$ be the
maximal time of existence and denote
\[
\tilde{\mathcal{P}}(t)=\sup_{0\leqslant s<t}\{|p|:f(s,x,p)\neq
0,\textnormal{ for some }x\in\R^3\}.
\]
In \cite{CR1, CR2} it is proved that
$\tilde{\mathcal{P}}(T_\mathrm{max})<\infty\Rightarrow
T_\mathrm{max}=\infty$, i.e., the solution could blow-up in finite
time only if the momentum support of $f$ becomes unbounded.
However, for the proof of Theorem \ref{main} it is essential to
look at another quantity. Define
\[
\mathcal{P}(t)=\sup_{0\leqslant
s<t}\{e^{\phi}\sqrt{1+|p|^2}:f(s,x,p)\neq 0,\textnormal{ for some
}x\in\R^3\},
\]
the maximal particles energy in the support of $f$.
\begin{Lemma}\label{equivalence}
The assertions $\tilde{\mathcal{P}}(t)<\infty$ and
$\mathcal{P}(t)<\infty$ are equivalent. In particular
\[
\mathcal{P}(T_\mathrm{max})<\infty\Rightarrow
T_\mathrm{max}=\infty.
\]
\end{Lemma}
\noindent\textit{Proof: }Since $e^\phi\leqslant C(t)$, then we
have $\mathcal{P}(t)\leqslant
C(t)\sqrt{1+\tilde{\mathcal{P}}(t)^2}$. Moreover, using
(\ref{estf})-(\ref{est1}),
\begin{equation}\label{estmu}
\mu(t,x)\leqslant\int_{|p|\leqslant
e^{-\phi}\mathcal{P}(t)}f\,\frac{dp}{\sqrt{1+|p|^2}}\leqslant
Ce^{2\phi}\mathcal{P}(t)^2\leqslant C(t)\mathcal{P}(t)^2;
\end{equation}
hence $\mathcal{P}(t)<\infty$ implies $\|\mu(t)\|_\infty\leqslant
C(t)$ and therefore also $\|\phi(t)\|_\infty\leqslant C(t)$. Thus
\[
\mathcal{P}(t)<\infty\Rightarrow\tilde{\mathcal{P}}(t)\leqslant
\left(\sup_{s\in [0,t)}
e^{\|\phi(s)\|_\infty}\right)\mathcal{P}(t)\leqslant C(t).
\]
\prfe In order to estimate the function $\mathcal{P}(t)$ we shall
use that, along characteristics,
\begin{equation}\label{crucial}
\frac{d}{ds}e^{2\phi}(1+|p|^2)=2e^{2\phi}\partial_s\phi.
\end{equation}
The aim is to transform (\ref{crucial}) in a Gr\"onwall's type
inequality by estimating $\partial_s\phi$ in terms of
$\mathcal{P}(t)$. An estimate like $|\partial_t\phi|\leqslant
C(t)\mathcal{P}(t)^2\log \mathcal{P}(t)$ would be enough. However
we are not able to obtain such a pointwise estimate for
$\partial_t\phi$. Rather we have to use the integral version of
(\ref{crucial}), namely
\begin{equation}\label{crucial2}
e^{2\phi}(1+|p|^2)=e^{2\phi_0(X(0))}(1+|P(0)|^2)+2\int_0^t
e^{2\phi}\partial_s\phi(s,X(s))\,ds.
\end{equation}
Eventually the quantity we shall estimate is the time integral in
the right hand side of (\ref{crucial2}). For this purpose we need
the integral representation formula for $\partial_t\phi$ which was
derived in \cite{CR1}. We recall it here for the sake of
reference:
\begin{equation}\label{reprdtphi1}
\partial_t\phi(t,x)=(\partial_t\phi)_D+\rm{I}+\rm{II}+\rm{III},
\end{equation}
where
\[
(\partial_t\phi)_D=\partial_t\phi_\mathrm{hom}-\frac{1}{t}\int_{|x-y|=t}\int\frac{f_0(y,p)}{(1+\omega\cdot\widehat{p})}\,\frac{dp}{\sqrt{1+|p|^2}}\,dS_y,
\]
\[
{\rm I}=\int_{|x-y|\leqslant t}\int\frac{(\omega+\widehat{p})\cdot\widehat{p}}{(1+\omega\cdot\widehat{p})^2}\,f(t-|x-y|,y,p)\,\frac{dp}{\sqrt{1+|p|^2}}\,\frac{dy}{|x-y|^2},
\]
\[
{\rm II}=-\int_{|x-y|\leqslant t}\int\frac{(\omega+\widehat{p})^2}{(1+\omega\cdot\widehat{p})^2}\,(S\phi)f(t-|x-y|,y,p)\,\frac{dp}{\sqrt{1+|p|^2}}\,\frac{dy}{|x-y|},
\]
\[
{\rm III}=-\int_{|x-y|\leqslant t}\int\frac{(\omega+\widehat{p})\cdot\nabla_x\phi}{(1+\omega\cdot\widehat{p})^2}\,f(t-|x-y|,y,p)\,\frac{dp}{(1+|p|^2)^{3/2}}\,\frac{dy}{|x-y|}.
\]
We rewrite the above representation formula in a new form which is
suitable for being estimated in terms of the null cone energy:
\begin{Proposition}\label{reprdtphi2}
The representation formula (\ref{reprdtphi1}) can be rewritten in the form
\[
\partial_t\phi(t,x)=(\partial_t\phi)_D+\sum_{i=0}^5\mathcal{Z}_i,
\]
where
\[
\mathcal{Z}_0=-2\int_{|x-y|\leqslant
t}(\partial_t\phi)\mu(t-|x-y|,y)\,\frac{dy}{|x-y|},
\]
\[
\mathcal{Z}_1=\int_{|x-y|\leqslant
t}\int\frac{f(t-|x-y|,y,p)}{\sqrt{1+|p|^2}(1+\omega\cdot\widehat{p})}\,dp\,\frac{dy}{|x-y|^2},
\]
\[
\mathcal{Z}_2=-\int_{|x-y|\leqslant
t}\int\frac{f(t-|x-y|,y,p)}{(1+|p|^2)^{3/2}(1+\omega\cdot\widehat{p})^2}\,dp\,\frac{dy}{|x-y|^2},
\]
\[
\mathcal{Z}_3=2\int_{|x-y|\leqslant
t}(\partial_t\phi-\omega\cdot\nabla_x\phi)\int\frac{(\omega\cdot\widehat{p})f(t-|x-y|,y,p)}{\sqrt{1+|p|^2}(1+\omega\cdot\widehat{p})}\,dp\,\frac{dy}{|x-y|},
\]
\[
\mathcal{Z}_4=\int_{|x-y|\leqslant
t}(\partial_t\phi-\omega\cdot\nabla_x\phi)\int\frac{f(t-|x-y|,y,p)}{(1+|p|^2)^{3/2}(1+\omega\cdot\widehat{p})^2}\,dp\,\frac{dy}{|x-y|},
\]
\[
\mathcal{Z}_5=-2\int_{|x-y|\leqslant
t}(\omega\wedge\nabla_x\phi)\cdot\int\frac{(\omega\wedge\widehat{p})f(t-|x-y|,y,p)}{\sqrt{1+|p|^2}(1+\omega\cdot\widehat{p})}\,dp\,\frac{dy}{|x-y|}.
\]
\end{Proposition}
\noindent\textit{Proof: }This proposition is the result of a
straightforward calculation which proceeds as follows. In the
integral ${\rm I}$ of (\ref{reprdtphi1}) we write
\[
(\omega+\widehat{p})\cdot\widehat{p}=(1+\omega\cdot\widehat{p})-(1+|p|^2)^{-1},
\]
which shows that ${\rm I}=\mathcal{Z}_1+\mathcal{Z}_2$. In the
integrals ${\rm II}$ and ${\rm III}$ of (\ref{reprdtphi1}) we
decompose $\nabla_x\phi$ into a component parallel to $\omega$ and
a component orthogonal to $\omega$, i.e., we write
\[
\nabla_x\phi=(\omega\cdot\nabla_x\phi)\omega-\omega\wedge\omega\wedge\nabla_x\phi.
\]
It follows that
\[
S\phi=(\partial_t\phi)(1+\omega\cdot\widehat{p})-(\partial_t\phi-\omega\cdot\nabla_x\phi)(\omega\cdot\widehat{p})+(\omega\wedge\nabla_x\phi)\cdot(\omega\wedge\widehat{p})
\]
and
\[
(\omega+\widehat{p})\cdot\nabla_x\phi=(\omega\cdot\nabla_x\phi)(1+\omega\cdot\widehat{p})+(\omega\wedge\nabla_x\phi)\cdot(\omega\wedge\widehat{p}).
\]
In the integral ${\rm II}$ we also use
\[
(\omega+\widehat{p})^2=2(1+\omega\cdot\widehat{p})-(1+|p|^2)^{-1}.
\]
After substituting and summing up the various terms one can easily
verify that
\[
{\rm II}+{\rm
III}=\mathcal{Z}_0+\mathcal{Z}_3+\mathcal{Z}_4+\mathcal{Z}_5,
\]
which concludes the proof.\prfe

We conclude this section with
\begin{Lemma}\label{calculus}
For $R>1$, $a,b\geqslant 0$, denote
\[
\mathcal{B}_{ab}(R)=\int_{|p|\leqslant
R}(\sqrt{1+|p|^2}+\omega\cdot p)^{-a}(1+|p|^2)^{-b}dp.
\]
Then
\begin{eqnarray*}
&&\mathcal{B}_{a,b}(R)\leqslant CR^{(2-2b)}\log R,\quad\textnormal{if }a=1,\,b<1;\\
&&\mathcal{B}_{a,b}(R)\leqslant CR^{(3-2b-a)},\quad\textnormal{if }a<1,\,b<\frac{3-a}{2};\\
&&\mathcal{B}_{a,b}(R)\leqslant CR^{(1+a-2b)},\quad\textnormal{if
}a>1,\,b<\frac{1+a}{2}.
\end{eqnarray*}
\end{Lemma}
\noindent\textit{Proof: }The proof is by evaluating the integral
in polar coordinates.\prfe
\section{Proof of the main theorem}
Without loss of generality, we can assume $\mathcal{P}(t)\geqslant
C$, where $C$ can be chosen arbitrarily large, otherwise we
redefine $\mathcal{P}(t)\rightarrow\mathcal{P}(t)+C$. A first
pointwise estimate on $\partial_t\phi$ follows by the results of
the previous section.
\begin{Proposition}\label{pointest}
\begin{eqnarray*}
|\partial_t\phi(t,x)|\leqslant2\int_{|x-y|\leqslant
t}|\partial_t\phi|\mu(t-|x-y|,y)\,\frac{dy}{|x-y|}
+C(t)\mathcal{P}(t)^2\log \mathcal{P}(t).
\end{eqnarray*}
\end{Proposition}
\noindent\textit{Proof: }From Proposition \ref{reprdtphi2} we
have
\begin{equation}\label{boh}
|\partial_t\phi(t,x)|\leqslant C(t)+2\int_{|x-y|\leqslant
t}|\partial_t\phi|\mu(t-|x-y|,y)\frac{dy}{|x-y|}+\sum_{i=1}^5|\mathcal{Z}_i|.
\end{equation}
Let us estimate each integral $\mathcal{Z}_i$, for $i=1,\dots 5$.
Observe that the domain of integration in the variable $p$ can be
chosen as $\{|p|\leqslant 1+e^{-\phi}\mathcal{P}(t)\}$ by the
definition of $\mathcal{P}(t)$.

\noindent\textit{Estimate for $\mathcal{Z}_1$}: By
(\ref{estf})--(\ref{est2}) and Lemma \ref{calculus},
\begin{eqnarray*}
|\mathcal{Z}_1|&\leqslant& C\int_{|x-y|\leqslant
t}e^{4\phi}\mathcal{B}_{1,0}\left(1+e^{-\phi}\mathcal{P}(t)\right)(t-|x-y|,y)\frac{dy}{|x-y|^2}\\
&\leqslant& C(t)\mathcal{P}(t)^2\int_{|x-y|\leqslant
t}e^{2\phi}\log\left(1+e^{-\phi}\mathcal{P}(t)\right)(t-|x-y|,y)\frac{dy}{|x-y|^2}\\
&\leqslant& C(t)\mathcal{P}(t)^2\int_{|x-y|\leqslant
t}\Big(e^{2\phi}|\phi|(t-|x-y|,y)+\log
\mathcal{P}(t)\Big)\frac{dy}{|x-y|^2}\\
&\leqslant& C(t)\mathcal{P}(t)^2\log \mathcal{P}(t).
\end{eqnarray*}

\noindent\textit{Estimate for $\mathcal{Z}_2$}: Again by
(\ref{estf})--(\ref{est2}) and Lemma \ref{calculus},
\begin{eqnarray*}
|\mathcal{Z}_2|&\leqslant& C\int_{|x-y|\leqslant
t}e^{4\phi}\mathcal{B}_{2,1/2}\left(1+e^{-\phi}\mathcal{P}(t)\right)(t-|x-y|,y)\frac{dy}{|x-y|^2}\\
&\leqslant& C(t)\mathcal{P}(t)^2\int_{|x-y|\leqslant
t}e^{2\phi(t-|x-y|,y)}\frac{dy}{|x-y|^2}\leqslant
C(t)\mathcal{P}(t)^2.
\end{eqnarray*}

\noindent\textit{Estimate for $\mathcal{Z}_3$}: By the
Cauchy-Schwarz inequality, (\ref{nullconeenergyestimate}),
(\ref{estf})--(\ref{est2}) and Lemma \ref{calculus},
\begin{eqnarray*}
|\mathcal{Z}_3|&\leqslant& C\int_{|x-y|\leqslant
t}|\partial_t\phi-\omega\cdot\nabla_x\phi|e^{4\phi}\mathcal{B}_{1,0}\left(1+e^{-\phi}\mathcal{P}(t)\right)(t-|x-y|,y)\frac{dy}{|x-y|}\\
&\leqslant&C\left(\int_{|x-y|\leqslant
t}|\partial_t\phi-\omega\cdot\nabla_x\phi|^2(t-|x-y|,y)\right)^{1/2}\\
&&\times \left(\int_{|x-y|\leqslant
t}e^{8\phi}\left[\mathcal{B}_{1,0}\left(1+e^{-\phi}\mathcal{P}(t)\right)\right]^2(t-|x-y|,y)\frac{dy}{|x-y|^2}\right)^{1/2}\\
&\leqslant& C(t)\mathcal{P}(t)^2\log \mathcal{P}(t).
\end{eqnarray*}

\noindent\textit{Estimate for $\mathcal{Z}_4$}: As before,
\begin{eqnarray*}
|\mathcal{Z}_4|&\leqslant& C\left(\int_{|x-y|\leqslant
t}|\partial_t\phi-\omega\cdot\nabla_x\phi|^2(t-|x-y|,y)dy\right)^{1/2}\\
&&\quad\quad\times\left(\int_{|x-y|\leqslant
t}e^{8\phi}\left[\mathcal{B}_{2,1/2}\left(1+e^{-\phi}\mathcal{P}(t)\right)\right]^2(t-|x-y|,y)\frac{dy}{|x-y|^2}\right)^{1/2}\\
&\leqslant& C(t)\mathcal{P}(t)^2.
\end{eqnarray*}

\noindent\textit{Estimate for $\mathcal{Z}_5$}: Again as before,
\begin{eqnarray*}
|\mathcal{Z}_5|&\leqslant& C\left(\int_{|x-y|\leqslant
t}|\omega\wedge\nabla_x\phi|^2(t-|x-y|,y,p)\,dy\right)^{1/2}\\
&&\times\left(\int_{|x-y|\leqslant
t}e^{8\phi}\left[\mathcal{B}_{1,0}\left(1+e^{-\phi}\mathcal{P}(t)\right)\right]^2(t-|x-y|,y)\frac{dy}{|x-y|^2}\right)^{1/2}\\
&\leqslant&C(t)\mathcal{P}(t)^2\log\mathcal{P}(t).
\end{eqnarray*}
Replacing the preceding estimates in (\ref{boh}) concludes the
proof.\prfe

Using Proposition \ref{pointest} and (\ref{crucial2}) we obtain
the integral inequality
\begin{eqnarray}
e^{2\phi}(1+|p|^2)&\leqslant& C+2\int_0^t
e^{2\phi}|\partial_s\phi(s,X(s))|\,ds \nonumber\\
&\leqslant&C+C(t)\int_0^t\mathcal{P}(s)^2\log
\mathcal{P}(s)\,ds+4I_0(|\partial_t\phi|\mu,t),\label{crucial3}
\end{eqnarray}
where
\[
I_0(g,t)=\int_0^te^{2\phi(s,X(s))}\int_{|X(s)-y|\leqslant
s}g(s-|X(s)-y|,y)\,\frac{dy}{|X(s)-y|}\,ds.
\]
We rewrite $I_0(g,t)$ as
\[
I_0(g,t)=\int_0^t\mathcal{I}_0(g,\tau,t)\,d\tau,
\]
where
\[
\mathcal{I}_0(g,\tau,t)=\int_\tau^{t}e^{2\phi(s,X(s))}\int_{|y|=s-\tau}\frac{g(\tau,X(s)-y)}{(s-\tau)}\,dS_y\,ds.
\]
Except for the factor $e^{2\phi(s,X(s))}$,
$\mathcal{I}_0(g,\tau,t)$ is the integral which is estimated in
the proof of \cite[Lemma~1.1]{Pa}. However, since we shall need a
slightly different formulation of the estimate proved in
\cite{Pa}, we present here a complete proof of the result that we
are going to use:
\begin{Lemma}\label{Pallard}
For all $0\leqslant \tau\leqslant t$,
\[
\mathcal{I}_0(g,\tau,t)\leqslant
C(t)\frac{\|g(\tau)\|_{L^2}}{\sqrt{t-\tau}}\int_\tau^{t}\log
\mathcal{P}(s)\,ds.
\]
\end{Lemma}
\noindent\textit{Proof:} We first rewrite $\mathcal{I}_0$ in
spherical coordinates:
\[
\mathcal{I}_0(g,\tau,t)=\int_\tau^{t}e^{2\phi(s,X(s))}\int_0^\pi\int_0^{2\pi}g(\tau,X(s)-(s-\tau)\omega)(s-\tau)\sin\theta\,d\varphi\,d\theta\,ds,
\]
where
$\omega=\omega(\theta,\varphi)=(\sin\theta\cos\varphi,\sin\theta\sin\phi,\cos\theta)$.
Now, in \cite[Lemma~2.1]{Pa} it is shown that the transformation
of variables $(s,\theta,\varphi)\rightarrow X(s)-(s-\tau)\omega$
is a $C^1$ diffeomorphism with Jacobian
\[
J=(\dot{X}(s)\cdot\omega-1)(s-\tau)^2\sin\theta=(\widehat{P}(s)\cdot\omega-1)(s-\tau)^2\sin\theta.
\]
Hence, applying Cauchy-Schwarz's inequality,
\begin{eqnarray*}
\mathcal{I}_0(g,\tau,t)&\leqslant&\left(\int_\tau^{t}\int_0^{\pi}\int_0^{2\pi}g^2(\tau,X(s)-(s-\tau)\omega)|J|\,d\varphi\,d\theta\,ds\right)^{1/2}\\
&&\quad\quad\times\left(\int_\tau^{t}e^{4\phi(s,X(s))}\int_0^\pi\int_0^{2\pi}\frac{\sin\theta}{(1-\widehat{P}(s)\cdot\omega)}\,d\varphi\,d\theta\,ds\right)^{1/2}\\
&\leqslant&\|g(\tau)\|_{L^2}\left(\int_\tau^{t}e^{4\phi(s,X(s))}\int_0^\pi\int_0^{2\pi}\frac{\sin\theta}{(1-\widehat{P}(s)\cdot\omega)}\,d\varphi\,d\theta\,ds\right)^{1/2}.
\end{eqnarray*}
We estimate the angular integral as
\begin{eqnarray*}
\int_0^\pi\int_0^{2\pi}\frac{\sin\theta}{(1-\widehat{P}(s)\cdot\omega)}\,d\varphi\,d\theta\,ds&=&2\pi\int_{-1}^1\frac{du}{(1-|\widehat{P}(s)|u)}\\
&\leqslant&C\left(1-\log(1-|\widehat{P}(s)|)\right)\leqslant
C\left(|\phi|+\log\mathcal{P}(s)\right).
\end{eqnarray*}
We finally obtain
\begin{eqnarray*}
\mathcal{I}_0(g,\tau,t)&\leqslant&
C(t)\|g(\tau)\|_{L^2}\left(\int_\tau^{t}\left(e^{\phi}|\phi|+\log\mathcal{P}(s)\right)ds\right)^{1/2}\\
&\leqslant&
C(t)\frac{\|g(\tau)\|_{L^2}}{\sqrt{t-\tau}}\int_\tau^{t}\log\mathcal{P}(s)\,ds,
\end{eqnarray*}
which concludes the proof of the lemma.\prfe

The proof of Theorem \ref{main} is now almost complete. Observe
that, by (\ref{estmu}) and (\ref{energyestimate}),
\[
\|\partial_t\phi\mu(\tau)\|_{L^2}\leqslant
\|\mu(\tau)\|_{\infty}\|\partial_t\phi(\tau)\|_{L^2}\leqslant
C(t)\mathcal{P}(\tau)^2,\quad\tau\leqslant t.
\]
Thus
\[
\mathcal{I}_0(|\partial_t\phi|\mu,\tau,t)\leqslant
C(t)\frac{\mathcal{P}(\tau)^2}{\sqrt{t-\tau}}\int_\tau^t\log\mathcal{P}(s)\,ds.
\]
Hence the integral $I_0(|\partial_t\phi|\mu,t)$ is bounded by
\begin{eqnarray*}
I_0(|\partial_t\phi|\mu,t)&\leqslant&C(t)\int_0^t\int_\tau^t\frac{\mathcal{P}(\tau)^2}{\sqrt{t-\tau}}\log\mathcal{P}(s)\,ds\,d\tau\\
&=&C(t)\int_0^t\int_0^s\frac{\mathcal{P}(\tau)^2}{\sqrt{t-\tau}}\log\mathcal{P}(s)\,d\tau\,ds\\
&\leqslant&C(t)\int_0^t\mathcal{P}(s)^2\log\mathcal{P}(s)\,ds.
\end{eqnarray*}
Finally, going back to (\ref{crucial3}) we obtain the Gr\"onwall
inequality
\[
\mathcal{P}(t)^2\leqslant C(t)\left(1+\int_0^t\mathcal{P}(s)^2\log
\mathcal{P}(s)\,dt\right),
\]
whence $\mathcal{P}(t)\leqslant C(t)$. By Lemma \ref{equivalence},
this completes the proof of the main theorem.

\bigskip
\noindent {\bf Acknowledgments:} The author acknowledges support
by the European HYKE network (contract HPRN-CT-2002-00282) and by
the project ``PDE and Harmonic Analysis'', sponsored by Research
Council of Norway (proj. no. 160192/V30).

\end{document}